\documentclass{elsart4-1}

\usepackage{graphicx}
\usepackage{epsfig}

\usepackage{amssymb}

\usepackage[english,francais]{babel}

\usepackage{array}
\newcolumntype{L}[1]{>{\raggedright\let\newline\\\arraybackslash\hspace{0pt}}m{#1}}
\newcolumntype{C}[1]{>{\centering\let\newline\\\arraybackslash\hspace{0pt}}m{#1}}
\newcolumntype{R}[1]{>{\raggedleft\let\newline\\\arraybackslash\hspace{0pt}}m{#1}}

\newtheorem{e-proposition}[theorem]{Proposition}

\newtheorem{e-definition}[theorem]{Definition\rm}

\renewcommand{\theequation}{\arabic{equation}}
\def\og{\leavevmode\raise.3ex\hbox{$\scriptscriptstyle\langle\!\langle$~}}
\def\fg{\leavevmode\raise.3ex\hbox{~$\!\scriptscriptstyle\,\rangle\!\rangle$}}

\begin{document}

\centerline{Astrophysics}
\begin{frontmatter}



\selectlanguage{english}
\title{The future of gamma-ray astronomy}


\selectlanguage{english}
\author{J\"urgen Kn\"odlseder}
\ead{jknodlseder@irap.omp.eu}

\address{IRAP, 9 avenue du Colonel Roche\\
31028 Toulouse Cedex 4, France}

\medskip
\begin{center}
{\small Received *****; accepted after revision +++++}
\end{center}

\begin{abstract}

The field of gamma-ray astronomy has experienced impressive progress over the last
decade.
Thanks to the advent of a new generation of imaging air Cherenkov telescopes
(H.E.S.S., MAGIC, VERITAS)
and thanks to the launch of the Fermi-LAT satellite, several thousand gamma-ray sources
are known today, revealing an unexpected ubiquity of particle acceleration processes in the
Universe.
Major scientific challenges are still ahead, such as the identification of the nature of Dark Matter,
the discovery and understanding of the sources of cosmic rays,
or the comprehension of the particle acceleration processes that are at work in the various objects.
This paper presents some of the instruments and mission concepts that will address these
challenges over the next decades.

{\it To cite this article: J. Kn\"odlseder, C. R. Physique {\bf TBD} (2015).}

\vskip 0.5\baselineskip

\selectlanguage{francais}
\noindent{\bf R\'esum\'e}
\vskip 0.5\baselineskip
\noindent

Le domaine de l'astronomie gamma a connu des progr\`es impressionnants au cours de la
derni\`ere d\'ecennie.
Gr\^ace \`a l'av\`enement d'une nouvelle g\'en\'eration de t\'elescopes Tcherenkov
(H.E.S.S., MAGIC, VERITAS)
et gr\^ace au lancement du satellite Fermi-LAT, plusieurs milliers de sources de rayons gamma
sont connus aujourd'hui, r\'ev\'elant une ubiquit\'e inattendue des processus d'acc\'el\'eration de
particules dans l'Univers.
Toutefois, des questions scientifiques majeures restent en suspens, telles que 
l'identification de la nature de la mati\`ere sombre,
la d\'ecouverte et la compr\'ehension des sources de rayons cosmiques,
ou la compr\'ehension des processus d'acc\'el\'eration de particules qui sont \`a l'oeuvre
dans les diff\'erents astres.
Cet article pr\'esente quelques-uns des instruments et des concepts de mission qui vont relever
ces d\'efis au cours des prochaines d\'ecennies.

{\it Pour citer cet article~: J. Kn\"odlseder, C. R. Physique {\bf TBD} (2015).}

\vskip 0.5\baselineskip
\noindent{\small{\it Keywords~:} Gamma rays; astronomy; dark matter; cosmic rays; particle acceleration
\vskip 0.5\baselineskip
\noindent{\small{\it Mots-cl\'es~:} Rayons gamma~; astronomie~; mati\`ere sombre~; rayons cosmiques~;
acc\'el\'eration de particules}}

\end{abstract}
\end{frontmatter}


\selectlanguage{english}

\section{Introduction}
\label{sec:intro}

Gamma-ray astronomy covers observations of photons with energies above a few 100~keV, with
current instruments reaching up to about 100~TeV.
Formally, there is no upper energy limit to gamma rays, yet pair-production on background
photons will effectively set a horizon to the explorable Universe.
For photon energies above $\sim1$~PeV this horizon is of the size of our Galaxy~\cite{gould1966}.
Gamma rays interact with the Earth's atmosphere, hence their direct detection from the terrestrial
surface is not possible.
Gamma rays are thus either observed directly from space using detectors aboard satellites or 
stratospheric balloons, or indirectly from ground by detecting the electromagnetic cascades
that are generated by gamma-ray induced pair production in the Earth atmosphere.
Gamma-ray instruments comprise
coded-mask telescopes for the low-energy range
(e.g.~INTEGRAL~\cite{winkler1994}),
Compton telescopes for the medium-energy range
(e.g.~COMPTEL~\cite{schonfelder93}),
pair creation telescopes for the high-energy range
(e.g.~Fermi~\cite{atwood2009}, AGILE~\cite{pittori2004}),
Cherenkov telescopes for the very-high-energy range 
(e.g.~H.E.S.S.~\cite{aharonian2006a}, MAGIC~\cite{aleksic2012}, VERITAS~\cite{holder2006},
MILAGRO~\cite{atkins2003}), and
charged particle detectors or integrating non-imaging Cherenkov detectors for the ultra-high-energy
range (e.g.~AIROBICC~\cite{aharonian2002}).
For a review of these detection techniques, see~\cite{SPACE,GROUND} and references therein.

Detection of the first celestial gamma-ray sources has been achieved
in the late 1950ies in the low-energy domain~\cite{peterson1958},
in the early 1960ies in the medium-energy~\cite{arnold1962} and high-energy
domains~\cite{kraushaar1962}, and in the late 1980ies in the very-high-energy 
domain~\cite{weekes1989}.
Since then, improvements in instrumental performances have differed between domains,
with the most spectacular results achieved so far in the high-energy range that today has
an inventory of over 3000 steady sources of gamma rays~\cite{acero2015}.
Progress has also been impressive in the very-high-energy domain, with well over 100
confirmed sources.
The low- and medium-energy gamma-ray domains have so far not experienced a 
comparable development, the number of few 100 keV and MeV steady sources being of the
order of several tens.
No gamma-ray source has so far been detected in the ultra-high-energy range, in the domain 
above $\sim100$~TeV.

The common feature of all gamma-ray sources is the non-thermal nature of the underlying
emission processes.
As opposed to thermal radiation that originates from the random movements of particles in 
matter with temperature above absolute zero, non-thermal radiation may have a variety of 
origins:
the decay or de-excitation of atomic nuclei,
the decay of particles or their annihilation with antiparticles,
and the interaction of non-thermal particle populations with photons and matter.
These processes may either lead to emission of mono-energetic photons
or to emission of a broad-band continuum spectrum of photons, covering eventually the
full electromagnetic spectrum from the radio band to the gamma-ray domain.
Within that spectrum, the gamma-ray band is unique since it is free from concurrent thermal
radiation that dilutes the non-thermal radiation at lower energies.
In other words, gamma rays provide the clearest view on the non-thermal physics in
our Universe, and for decay, de-excitation and annihilation processes they often provide 
the only view.

Gamma-ray astronomy is thus the astronomy of the non-thermal Universe.
Since its advent about 50 years ago it has revealed an unexpected variety of objects
that release a significant, sometimes even dominant fraction of their energy through
non-thermal processes, including amongst others neutron stars, black holes, stellar
explosions and their remnants.
The exploratory phase is certainly concluded for the high-energy and very-high-energy
range, which now will turn into mature fields of contemporary astronomy that will deepen
our understanding of the underlying physics.
For the low- and medium-energy domains, exploration has only started, and an important 
step needs to be taken to bring the non-thermal phenomenologies into light.
The ultra-high-energy domain is still terra incognita, but instruments exist that could soon
reveal first sources, provided that some exist within the accessible Universe~\cite{aartsen2013}.

This review aims to depict the evolution of the field of gamma-ray astronomy in the 
foreseeable future, based on recent achievements, open science questions, and ongoing
instrument developments.
The review will concentrate on the high-energy and very-high-energy range, which currently
are the most vital domains.
Specifically, the review will discuss future instruments (or instrument concepts) that are able
to detect gamma rays in the 100~MeV to 100~TeV energy range, although some of the 
discussed instruments will explore the sky beyond this band.

\section{Scientific challenges}
\label{sec:challenges}

\subsection{Dark matter}

The nature of dark matter is certainly one of the most fundamental problems of modern science.
Evidenced as apparently missing mass at scales of galaxies, galaxy clusters, but also the entire 
Universe, it inevitably points us to a flaw in our understanding of nature.
Proposed solutions to this mystery comprise modifications to the fundamental laws of
physics, as well as the introduction of new, weakly interacting particles that so far escaped
detection.
The most popular candidate for the latter are weakly interacting massive particles (WIMP)
that arise in extensions to the standard model of particle physics~\cite{bertone2005}.
Direct searches for WIMPs are currently performed in underground laboratories and 
using the Large Hadron Collider, while indirect searches rely on the detection of annihilation 
and decay products that may lead to observable signatures in the gamma-ray domain~\cite{COSMO}.
Less popular but not less interesting candidates comprise the axion that may leave an
imprint on gamma-ray spectra of distant sources~\cite{PROBES}.

Annihilation or decay of dark matter particles is expected to lead to gamma-ray continuum and
line emission in the GeV -- TeV domain~\cite{bertone2005}.
The most stringent limits today come from the Fermi-LAT telescope which has ruled out the
existence of WIMP particles with masses $<30$~GeV~\cite{ackermann2011a}.
These measurements are corroborated by recent upper limits obtained with the Planck satellite
on the maximum WIMP annihilation cross section in the early Universe~\cite{galli2014}.
At higher energies, existing upper limits are less constraining, and current measurements
cannot exclude the existence of WIMP particles with masses above a few 10~GeV.
Next generation very-high-energy telescopes will be decisive to probe WIMP particles with
higher masses.
Any detection would mark a historical scientific breakthrough, while an upper limit would put
scientific constraints that will question the WIMP interpretation of dark matter.
Whatever the outcome, future very-high-energy gamma-ray telescopes will change our 
understanding of the Universe.

\subsection{Cosmic rays}

Since the discovery of cosmic rays at the beginning of the last century~\cite{hess1911}, it has
been established that the Earth is immersed in a basically isotropic flow of high-energy particles,
composed primarily of atomic nuclei and traces of electrons, positrons and antiprotons.
Particle energies span from a few 100~MeV to beyond $10^{20}$~eV, with a differential
energy spectrum following a power law that steepens at a few PeV before flattening at about an 
EeV and then turning over and terminating at a few 100~EeV~\cite{mello2014}.
It is widely accepted that particles up to the ``knee'' at a few PeV originate from sources within 
our Galaxy, with the ``knee'' marking the end of the Galactic proton spectrum~\cite{amato2014}.
There is rather convincing and yet circumstantial evidence that the bulk of the Galactic
cosmic rays are accelerated in supernova remnants (SNRs), though no proof has been found yet
that SNRs can accelerate cosmic rays up to the ``knee'' energy~\cite{SNRPWN}.
The acceleration of cosmic-ray protons and nuclei can be probed through observations of
gamma rays resulting from the decay of secondary $\pi^0$-mesons produced in hadronic
interactions, with the gamma-ray energy being roughly $10\%$ of the kinetic proton energy.
Detecting gamma rays up to 100~TeV from a SNR would unambiguously establish hadronic
cosmic-ray acceleration to $\sim$PeV energies since alternative emission channels, such
as inverse Compton scattering of high-energy leptons, are not believed to reach such high
energies~\cite{aharonian2013}.
Discovering such a Galactic ``PeVatron'' is without doubt the holy grail of very-high-energy
gamma-ray astronomy.
The enabling requirements for such a discovery are an improvement of the instrument
sensitivity in the energy-range up to $\sim100$~TeV, a domain that is only poorly covered
by existing instruments.

At low energies (below a few GeV), cosmic rays play a crucial role in the energetic balance of 
the interstellar medium and they are the primary driving agent of interstellar chemistry that ultimately 
produces the building blocks of life~\cite{viti2012}.
Low-energy cosmic rays are however heavily affected by the solar wind, hence their spectrum
and intensity escapes direct measurements.
Information can be indirectly inferred  through ionisation rate measurements of the interstellar
medium, suggesting for example an enhanced cosmic-ray ionisation rate in the vicinity of
gamma-ray producing supernova remnants~\cite{vaupre2014}.
Gamma rays can also be used directly to determine the spectrum of cosmic rays in interstellar
space~\cite{aharonian2006b,dermer2013}, yet current instruments are lacking sufficient 
low-energy sensitivity (below a few $100$~MeV) to provide stringent results.
Indirect measurements can be extended to lower cosmic-ray energies by observing  
gamma-ray lines in the MeV domain that result from cosmic-ray induced excitation of atomic
nuclei in the interstellar medium.
So far, these lines escape detection due to a lack in sensitivity of current instrumentation, but
they may be in reach of future medium-energy gamma-ray telescopes~\cite{benhabiles2013}.
In summary, a next generation medium- to high-energy gamma-ray satellite should enable the
exploration of the low-energy cosmic-ray component, shedding thus light on a crucial agent
that is key to Galactic physical and chemical processes.

Gamma rays not only carry information about the cosmic-ray intensity and spectrum, but also
about the propagation physics.
Specifically, if the mass of a molecular cloud is known, gamma-ray observations can be used
to measure the cosmic-ray energy density and to infer its spatial variations due to cosmic-ray
propagation~\cite{casanova2010}.
For example, observations of TeV gamma-ray emission from molecular clouds in the vicinity
of SNR~W28 suggest a reduced cosmic-ray diffusion coefficient with respect
to the average Galactic value, possibly due to the higher plasma turbulence around the
remnant~\cite{gabici2010}.
Models predict a concave-shaped gamma-ray spectrum as result of the superposition of background
cosmic rays with freshly accelerated nuclei that carries information about the diffusion
properties of the interstellar medium~\cite{gabici2009}.
Current instruments are not sensitive enough to reveal the predicted signatures, but future
very-high-energy gamma-ray facilities with improved sensitivity should allow inferring the
properties of cosmic-ray propagation from the observations.

Cosmic-ray phenomena are not limited to the Milky Way, as illustrated by the observation
of synchrotron emission from external galaxies attributed to cosmic-ray electrons
interacting with magnetic fields~\cite{condon1992}.
More recently, high-energy and very-high-energy gamma-ray emission has been observed 
from both local group and nearby starburst galaxies, which has been interpreted as 
cosmic-ray induced $\pi^0$-decay radiation~\cite{STARBURST}.
Gamma-ray observations of external galaxies provide the potential to study cosmic-ray
acceleration and transport in various global conditions~\cite{martin2014}, yet current
instruments lack the sensitivity to clearly distinguish between the cosmic-ray induced
and concurrent potential source components~\cite{ohm2013}.
Furthermore, the gamma-ray emission is generally not resolved, and even where spatial
information is available (as for example for the Magellanic Clouds), the limited angular resolutions 
of current instruments severely hampers the identification of cosmic-ray sources~\cite{abdo2010}.
An improved sensitivity combined with an enhanced angular resolution is therefore mandatory
to exploit the full potential of gamma-ray observations of external star-forming galaxies, and
to learn about cosmic-ray physics in different galactic environments.

\subsection{Particle acceleration}

As testified by current gamma-ray observations, particle acceleration is a ubiquitous
phenomenon in the Universe.
Although the basic physical processes that can lead to particle acceleration are understood,
the actual mechanism at operation in a specific source or the nature of the underlying source
is often poorly known~\cite{blandford2014}.
For example, the source of the highest energy cosmic rays is still elusive, and proposed
origins include
active galactic nuclei (AGN)~\cite{berezhko2008},
gamma-ray bursts (GRBs)~\cite{GRB},
spinning magnetars~\cite{fang2012},
and intergalactic shock fronts~\cite{kang1996}.
It is also not clear whether supernova remnant shocks are at the origin of particles of the
cosmic-ray spectrum between the "knee" and the "ankle", or whether larger shocks associated
with supershells or a Galactic wind termination shock are needed~\cite{blandford2014}.
So far, only little evidence for particle acceleration in supershells exist, although
recent high-energy~\cite{ackermann2011b} and very-high-energy~\cite{bartoli2014}
gamma-ray observations of the Cygnus~X region show evidence for hadronic particle 
acceleration in the Cygnus superbubble.
To confirm this scenario, sensitive $\sim100$~MeV observations are needed to reveal the
characteristic $\pi^0$ signature of hadronic emission.
And deeper very-high-energy observations should clarify whether Cygnus~X remains an
exception, or whether other particle accelerating supershells exist within our Galaxy
or its neighbours~\cite{abramowski2015}.

New insights into the physics of particle acceleration have recently been gained through
the study of gamma-ray variability.
For example, AGILE and Fermi-LAT observations have revealed dramatic few hours long
gamma-ray flares at a few 100 MeV from the Crab pulsar wind nebula that apparently have
no counterpart in any other wavelength band (see~\cite{buehler2014} for a review).
The observations indicate very efficient particle acceleration going beyond the classical
reaction limit in an MHD setting~\cite{blandford2014}.
Extremely rapid gamma-ray variability has also been observed from blazars on timescales
that can be as short as few minutes~\cite{AGN}.
In the most extreme cases, causality constrains the size of the emission region to a
fraction of the gravitational radius of the central black hole, being difficult to reconcile with
the conventional model of particle acceleration at internal shocks in the jet which should
act (at least) at dimensions of the black hole horizon~\cite{aleksic2014}.
The problem can be settled by accessing sub-minute timescales, which can be reached by
very-high-energy gamma-ray telescopes with an increased detection area compared to
existing instruments.
With these upcoming observations, time-domain astronomy will get elevated to becoming a 
major contributor to our understanding of particle acceleration processes in the
Universe~\cite{sanders2013}.
This is valid for the very-high-energy domain, but also for the few 100~MeV domain where the
end of the synchrotron spectrum of the most powerful particle accelerators can be
probed.

Among known particle acceleration sites, a few are still eluding detection in gamma rays.
This comprises particle acceleration in shocks generated by stellar winds interacting with the
interstellar medium~\footnote{On the other hand, the shocks generated by colliding winds in
binary systems~\cite{BINARIES} have been shown to accelerate particles and produce
gamma rays.} and through merger and accretion shocks as well as turbulences within galaxy
clusters.
The identification of synchrotron radio emission from a few massive stars provides evidence
that some of them are able to accelerate particles up to relativistic energies~\cite{debecker2013}.
Diffuse radio synchrotron emission is also observed from several galaxy clusters indicating
the presence of highly relativistic particles and large-scale magnetic fields~\cite{feretti2012}.
Whether both environments are also proton accelerators, and hence potential contributors
to the cosmic-ray particle population, is still an outstanding question.~\footnote{Recent Fermi-LAT
observations of gamma-ray emission towards $\eta$~Carinae have been interpreted as evidence
for proton acceleration~\cite{farnier2011}, yet a convincing detection of the expected orbital
modulation is still lacking~\cite{reitberger2012}.}
As for other source classes, gamma-ray observations have the unique potential to establish
these sources as proton accelerators.
An evident requirement for this is a good sensitivity in the few 100~MeV domain up to TeV
energies.
Angular resolution is also key, specifically for the stellar wind scenario, because potential
source regions are crowded with particle accelerators (pulsars, pulsar wind nebulae, 
gamma-ray binaries, supernova remnants).
An unambiguous spatial association between a gamma-ray source and a known non-thermal
radio emitting massive star is fundamental before any firm detection can be claimed.

\subsection{Terrae incognitae}

In addition to the aforementioned science challenges, there still exist {\it Terrae incognitae} in the 
gamma-ray domain, specifically in the still poorly explored low- to medium-energy range, as well 
as in the ultra-high-energy range.
Naturally, any instrument with improved performance has the potential to uncover the unexpected,
as has recently been demonstrated in the high-energy and very-high-energy domains.
Few have expected the ubiquity of gamma-ray emission in the Universe revealed at very high
energies by H.E.S.S., MAGIC, VERITAS and MILAGRO.
Almost nobody has envisioned classical novae as high-energy gamma-ray sources that
give a view on the dynamics of diffusive shock acceleration of protons in our Galaxy, as
has recently been demonstrated by Fermi-LAT (yet see~\cite{tatischeff2007}).
And the discovery of the Fermi bubbles came probably as a surprise to everyone~\cite{GC}.
These are a few selected examples that illustrate how future gamma-ray instruments will
push our understanding beyond the established horizons.
The remainder of this paper will now show how these future gamma-ray instruments will look like.

\section{Space-based gamma-ray astronomy}
\label{sec:space}

\subsection{The next steps}

Space-based gamma-ray astronomy basically covers the energy range below a few tens of GeV,
i.e. the low, medium- and high-energy gamma-ray domains.
In the low- and medium-energy domains, detection sensitivities are still relatively poor, yet the
technologies exist to build an affordable instrument that goes well beyond the sensitivity of the
INTEGRAL satellite or the COMPTEL telescope.
For example, modern and space-proven highly pixelised semiconductors alike those flying
currently on AGILE or Fermi-LAT~\cite{SPACE} can be arranged in a compact configuration with a
minimum amount of passive material (and in particular without conversion foils) to allow the
measurement of gamma-ray photons through both Compton and pair production interactions.
This would lead to a system that is sensitive from the MeV to the GeV domain, going about
more than a factor 10 beyond predecessor instruments, opening up this poorly explored energy
band.
In addition, such a system would be sensitive to the polarisation of incoming gamma rays,
accessing a new physical dimension that provides invaluable information about the underlying 
emission processes and source geometries.

At higher energies, succeeding to Fermi-LAT will be challenging.
The Fermi-LAT tracker has a geometric area of $1.5\times1.5$~m$^2$, the spacecraft has a diameter
of 2.5~m, a height of 2.8~m and a mass of 4.3 tons, which is a very respectable satellite.
Making the telescope much bigger to increase the effective detection area would soon hit the
maximum capacities of existing (and also planned) launch vehicles.
So the detection area can not be substantially expanded.
Significant improvement in sensitivity for pair-creation telescopes can only be achieved through
a dramatic improvement in the angular resolution, especially at lower energies where the 
Fermi-LAT point spread function (PSF) exceeds several degrees.
The best ways to improve the PSF are to decrease the density of the material in the tracker and
to space the tracking element further apart~\cite{charles2014}.
This joins pretty much the needs expressed above for the low- to medium-energy domains,
and both needs can be combined in a single mission providing a low-energy extension to
Fermi-LAT.

The following sections will describe some of the proposed instruments and mission concepts
for the future study of the gamma-ray Universe from space.
Table~\ref{tab:space} provides a summary of the performances of the proposed instruments
and mission concepts, as presented in the relevant references.
The accuracy of the performance predictions varies between instruments and also depends
on the maturity of the project, hence the table is only useful for order-of-magnitude comparisons.
The table also indicates the context (e.g.~``M4?'' indicates a candidate mission to ESA's medium-size
mission call M4) or the lead country and quoted or plausible launch dates.
Performance values are given for a reference energy, which is a ``typical'' energy in the core
energy range of the mission.
And ``t.b.d.'' indicates that the value has not been communicated or still needs to be defined.
Details are given in the relevant sections.

\begin{table}[!t]
 \centering
  \begin{tabular}{l|C{1.7cm}|C{1.7cm}|C{1.7cm}|C{1.7cm}|C{1.7cm}|C{1.7cm}|C{1.7cm}|C{1.7cm}}
   \hline \hline
   Parameter & AdEPT & e-ASTROGAM & CALET & DAMPE & GAMMA-400 & HARPO & HERD & PANGU \\
   \hline
   Context &
     R\&D & 
     M5? & 
     ISS & 
     China & 
     Russia & 
     R\&D & 
     China & 
     ESA/CAS? \\ 
   Launch date &
     -- & 
     2029? & 
     2015 & 
     2015 & 
     $\sim2021$ & 
     -- & 
     $>2020$ & 
     2021? \\ 
   Energy range (GeV) &
     0.005 - 0.2 & 
     0.0003 - 3 & 
     0.02 - 10000 & 
     2 - 10000 & 
     0.1 - 3000 & 
     0.003 - 3 & 
     0.1 - 10000 & 
     0.01 - 5 \\ 
   Ref.~energy (GeV) &
     $0.07$ & 
     $0.1$ & 
     $100$ & 
     $100$ & 
     $100$ & 
     $0.1$ & 
     $100$ & 
     $1$ \\ 
   $\Delta E/E$ &
     $30\%$ & 
     $30\%$ & 
     $2\%$ & 
     $1.5\%$ & 
     $1\%$ & 
     $10\%$ & 
     $1\%$ & 
     $30\%$ \\ 
   A$_{\rm eff}$ (cm$^2$) &
     500 & 
     1500 & 
     t.b.d. & 
     3000 & 
     5000 & 
     2700 & 
     t.b.d. & 
     180 \\ 
   Sensitivity (mCrab) &
     10 & 
     10 & 
     1000 & 
     100 & 
     100 & 
     1 & 
     10 & 
     t.b.d. \\ 
   Field of view (sr) & 
     t.b.d. & 
     $2.5$ & 
     $1.8$ & 
     $2.8$ & 
     $1.2$ & 
     t.b.d. & 
     t.b.d. & 
     $2.2$ \\ 
   Angular resolution &
     $1^\circ$ & 
     $1.5^\circ$ & 
     $0.1^\circ$ & 
     $0.1^\circ$ & 
     $0.02^\circ$ & 
     $0.4^\circ$ & 
     $0.1^\circ$ & 
     $0.2^\circ$ \\ 
   MDP (10 mCrab) &
     $10\%$ & 
     $20\%$ & 
     -- & 
     -- & 
     -- & 
     t.b.d. & 
     -- & 
     t.b.d. \\ 
   Technology &
     TPC & 
     Si+CsI & 
     fib.+PbWO$_4$ & 
     Si+BGO & 
     Si+CsI & 
     TPC & 
     Si+LYSO & 
     Si (fib.)+{\bf B}  \\ 
   \hline
  \end{tabular}
 \caption{Summary of instruments and mission concepts for space-based gamma-ray astronomy
 (see text). MDP indicates the minimum detectable polarisation of an instrument.
 Detector technologies comprise
 time projection chambers (TPC),
 silicon trackers (Si),
 cesium iodide scintillators (CsI),
 scintillating fibers (fib.),
 lead tungstate scintillators (PbWO$_4$),
 bismuth germanate scintillators (BGO),
 lutetium yttrium orthosilicate scintillators (LYSO), and
 magnetic spectrometers ({\bf B}).}
 \label{tab:space}
\end{table}

\subsection{AdEPT}

The {\it Advanced Energetic Pair Telescope} (AdEPT) targets the $5-200$ MeV energy range
and provides imaging, spectroscopy and polarimetry capabilities~\cite{hunter2014}.
The central element of AdEPT is a gaseous time projection chamber (TPC) filled with Ar+CS$_2$
at a pressure of 1.5 bar at $25^\circ$C.
The TPC serves as three-dimensional imager of electron-positron pair tracks arising from pair 
conversion of incoming photons in the detector volume.
The TPC has a dimension of $200\times200\times100$~cm$^3$ and is bounded by a drift electrode
on the top and a two-dimensional readout plane at the bottom, consisting of a micro-well detector
with a 400 $\mu$m pitch.
Signal pre-amplification is provided by a gas electron multiplier.
A field-shaping cage of wires creates a linear potential gradient of about 1 kV/cm within
the volume, the time of arrival of the signals on the readout plane enables the 3-D location
of the ionization charge in the volume.

Prototype detectors of volumes $10\times10\times15$~cm$^3$ and
$30\times30\times15$~cm$^3$ have been realized, and a
$30\times30\times7$~cm$^3$ detector was used for demonstration of neutron imaging in an
over-water environment~\cite{son2010}.
Building of a larger detector volume of $50\times50\times100$~cm$^3$ is planned.
Accelerator calibration measurements are foreseen to study gamma-ray polarisation
and to validate expected instrument performances.
A stratospheric balloon flight of a prototype detector is envisioned in the 2018-2020 time 
frame to validate the gamma-ray detection and background reduction capabilities.

Based on this R\&D work, a baseline concept for the AdEPT instrument and spacecraft
has been developed.
The AdEPT instrument would have a mass of 730~kg and a power consumption of 500~W.
The spacecraft would be 3-axis stabilised and zenith pointed, placed at a low altitude orbit 
(550 km) with $28^\circ$ inclination to assure a low cosmic-ray background.
The AdEPT sensitivity is estimated to be superior to the Fermi-LAT sensitivity below 
$\sim200$ MeV.
At the same time, the reduced Coulomb losses in the detector volume will bring a 5 fold 
improvement in angular resolution below $\sim200$ MeV compared with Fermi-LAT.
The minimum detectable polarisation (MDP) for a 10 mCrab~\footnote{In high-energy astrophysics,
the Crab nebula, one of the strongest steady sources, is often used as a standard candle.
The mCrab is thus the energy flux of a hypothetical source whose spectrum is scaled down from
that of the Crab nebula by a factor 1000.} source observed during $10^6$ seconds is estimated 
to $10\%$.

\subsection{e-ASTROGAM}

\begin{figure}[!t]
  \centering
  \includegraphics[height=8cm]{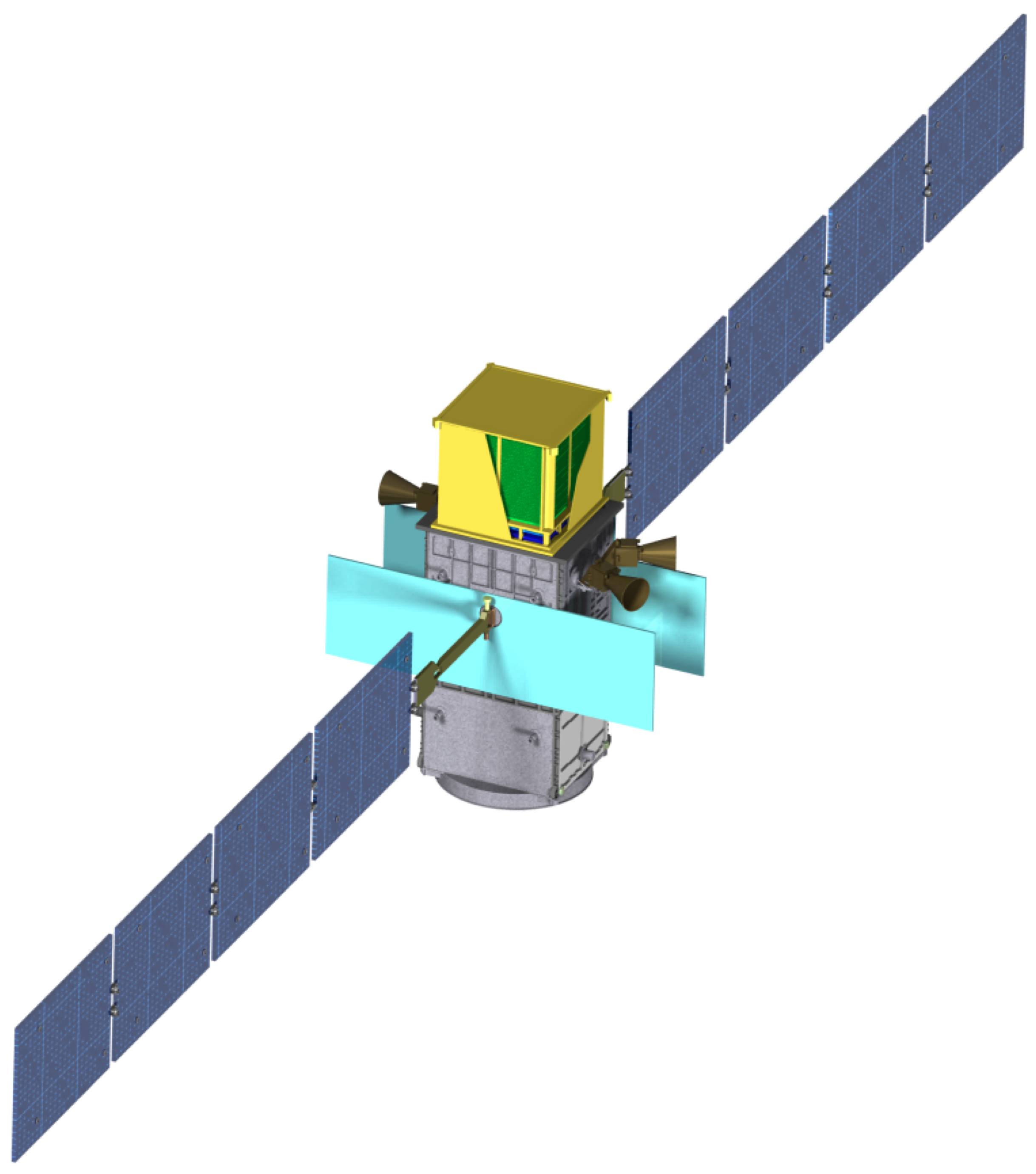}
  \caption{Artists view of the ASTROGAM satellite.}
  \label{fig:astrogam}
\end{figure}

The ASTROGAM mission concept~\cite{astrogam} has been proposed to ESA in the context of the
M4 Call for Missions, but unfortunately has not been selected for a mission study.
An extended version of the mission concept, e-ASTROGAM, is currently under preparation in
response to the M5 Call for Missions.
The e-ASTROGAM mission is an international project that targets the study of the low- to
medium-energy gamma-ray domain ($0.3$~MeV -- 3~GeV) by combining a Compton telescope
and a pair-conversion telescope into a single detector system~\cite{tatischeff2015}.
e-ASTROGAM builds, amongst others, on the experience of the Fermi-LAT and AGILE detector
developments.
By carefully minimising any passive material in the detector volume, as well as by omitting 
converter materials, e-ASTROGAM will provide unprecedented MeV to GeV sensitivity as well as 
polarisation measurement capabilities.

The e-ASTROGAM detector is composed of
a silicon tracker,
a 3D-imaging scintillator calorimeter, and
a plastic scintillator anti-coincidence shield.
In the configuration currently under study, the silicon tracker has a geometric surface area
of $95\times95$~cm$^2$ and is composed of 56 layers of $10\times10$ double-sided silicon
strip detectors (DSSDs) with a thickness of 500~$\mu$m, a strip pitch of 240~$\mu$m
and an inter-layer spacing of $7.5$~mm.
Each DSSD has a geometric area of $9.5\times9.5$~cm$^2$ with 384 strips in x and
y direction, respectively.
The DSSDs in each layer are wire bonded strip to strip to form $5\times5$ 2D ladders
that are readout by dedicated ASICs.
The total thickness of the tracker is 0.3 radiation lengths.
The calorimeter is made of 8\,464 CsI~(Tl) scintillator bars of 8~cm length (corresponding to 4.3
radiation lengths) and $10\times10$~mm$^2$ cross section.
The scintillation light of each bar is collected by $4\times4$ silicon drift diodes (SDDs) that are
glued at both ends to the crystals.
The SDDs are readout by dedicated ASICs.

The e-ASTROGAM detector has an overall volume of $110\times110\times60$~cm$^3$ and
a mass of 770~kg~\cite{tatischeff2015}.
It is proposed to use the European VEGA rocket or the Soyuz rocket to put e-ASTROGAM into
a low Earth orbit ($550-600$~km) with a low inclination angle of $<2.5^\circ$ that avoids the
high radiation environment of the South Atlantic Anomaly.
The satellite (cf.~Figure \ref{fig:astrogam}) will be 3-axis stabilised, and the observing strategy
will combine zenith-pointing sky-scanning observations with dedicated source pointings.
The nominal mission lifetime will be 3.5 years with a provision for a $>2$ years extension.
Table~\ref{tab:space} gives the performance value for a typical photon energy of 100~MeV.

The e-ASTROGAM observing time will be open to the international community through a Guest
Observer Programme.
A Core Science Program will be implemented to guarantee that the mission key objectives are
met.
All data will become public after 1 year of proprietary right.
Guest Observers will be assisted by the e-ASTROGAM Science Data Center with data and analysis
software.

\subsection{CALET}

The {\it CALorimetric Electron Telescope} (CALET, see~\cite{calet}) is a 
space-based detector developed by a Japanese-led international collaboration to directly 
measure the high-energy cosmic radiation on the International Space Station 
(ISS)~\cite{maestro2013}.
While CALET is primarily a particle detector, it will also be sensitive to gamma rays in the
20 MeV -- 10 TeV energy range.
CALET has been delivered to the ISS on 24 August 2015.

CALET consists of the charge detector (CHD), the finely segmented pre-shower imaging
calorimeter (IMC), and the total absorption calorimeter (TASC).
An incoming gamma ray pair converts in one of the 7 tungsten plates of the IMC
that are interleaved with orthogonal double layers of 1 mm$^2$ scintillating fibers.
The surface area of the IMC is $45\times45$ cm$^2$ and its total thickness is
3 radiation lengths.
The electron-positron pair then propagates into the TASC which is a homogeneous 
calorimeter made of 192 lead tungstate (PbWO$_4$) scintillator crystals arranged in
12 layers.
The TASC has a thickness of 27 radiation lengths and can determine the energy of the 
incident gamma-ray with an energy resolution of $\sim2\%$ (above 100 GeV).
Moreover, exploiting its shower imaging capabilities, a proton rejection $>10^5$ can be
achieved.

CALET has a mass of 480~kg and a power consumption of 310~W.
Mounted as a fixed payload on the Japanese Experiment Module Exposed Facility,
CALET continuously scan the sky as the ISS orbits Earth.
Although CALET cannot compete with the gamma-ray sensitivity of Fermi-LAT,
its excellent energy resolution provides a niche for detecting spectral line features,
such as those expected from dark matter particle decays~\cite{moiseev2013}.
Furthermore, CALET is equipped with a dedicated scintillator-based gamma-ray
burst monitor covering the 7 keV -- 20 MeV energy range that enables the detection of 
gamma-ray bursts and X-ray transients.

\subsection{DAMPE}

The {\it Dark Matter Particle Explorer} satellite (DAMPE, previously named TANSUO, 
see~\cite{dampe}) is a Chinese-led satellite with a detector unit that is conceptually 
very similar to that of CALET~\cite{wu2013}.
The DAMPE detector is primarily a particle detector with sensitivity in the gamma-ray
energy range from 2 GeV -- 10 TeV.

The DAMPE detector is composed of 
a plastic scintillator detector (PSD),
a 12-layer silicon tungsten tracker (STK),
a 14-layer bismuth germinate (BGO) calorimeter, and 
a neutron detector (ND).
For gamma-ray measurements, the PSD serves as anti-coincidence system for particle background
rejection.
Incoming photons pair-convert in the 3 tungsten plates of the STK that has a geometric area of 
$76\times76$ cm$^2$ and a thickness of $1.4$ radiation lengths.
The energy of the electron-positron pair is then absorbed in the BGO calorimeter that is composed
of 308 crystals read out by photomultiplier tubes.
The calorimeter has a thickness of 31 radiation lengths and determines the energy of the incident
gamma-ray with a resolution of $\sim1.5\%$ (above 100 GeV).

A prototype detector comprised of PSD, BGO and ND has been built and beam-tested at 
CERN to validate the detector performances~\cite{wu2013}.
The construction of the satellite hardware is in progress. 
DAMPE has a payload mass of 1480~kg and a power consumption of 500~W.
The satellite will be placed into a sun-synchronous low altitude orbit (500~km) with an inclination
of $97.4^\circ$.
At the time of writing this article, the launch with a Chinese Long March 2D rocket is scheduled 
for late 2015, the expected satellite lifetime is $5$~years.
Similar to CALET, the science niche of DAMPE is the detection of narrow gamma-ray lines
that may arise from the decay of dark matter particles~\cite{li2012}.

\subsection{GAMMA-400}

\begin{figure}[!t]
  \centering
  \includegraphics[width=8cm]{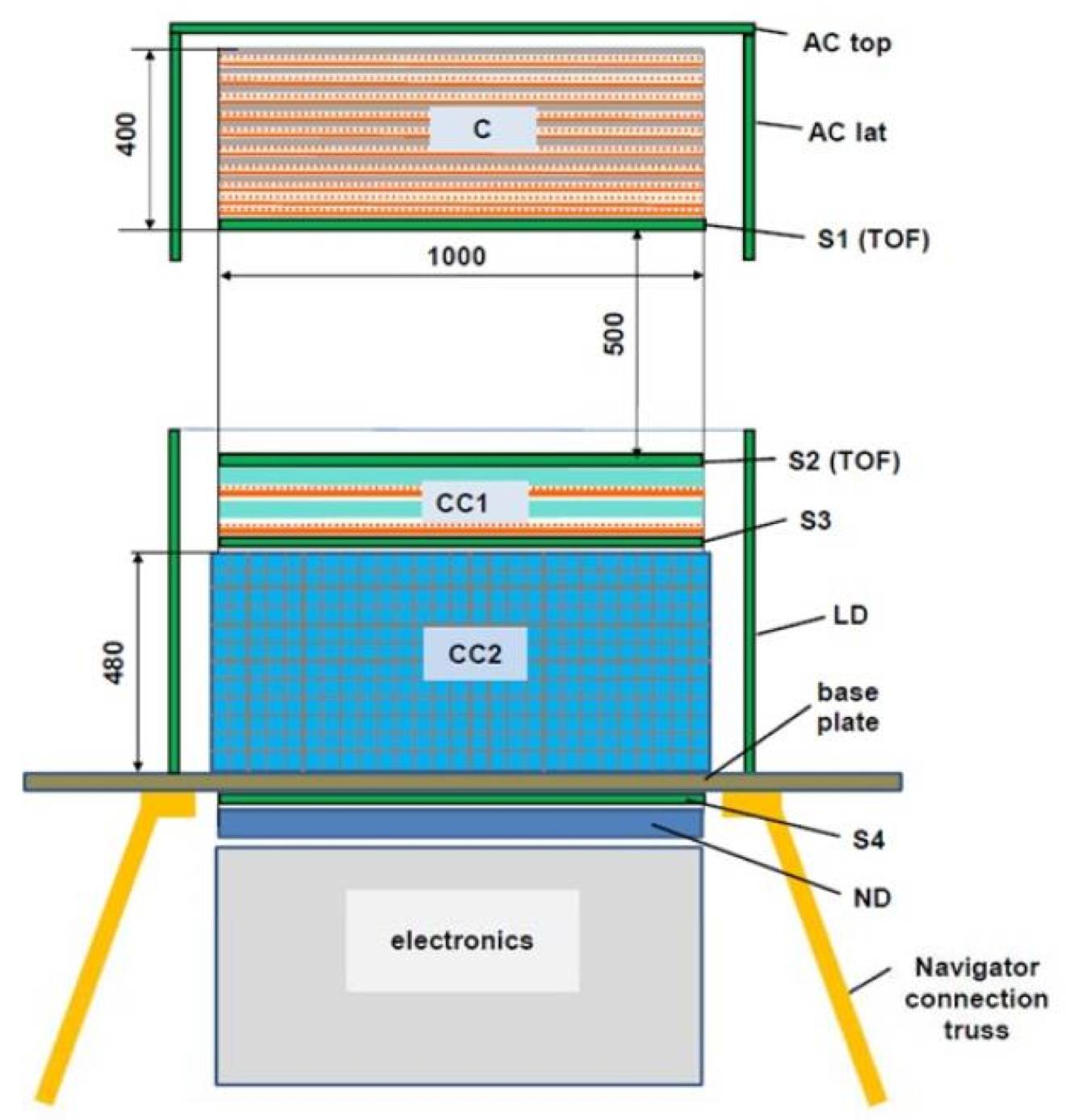}
  \caption{Concept of the Gamma-400 detector.}
  \label{fig:gamma400}
\end{figure}

The {\it Gamma Astronomical Multifunctional Modular Apparatus with the maximum gamma-ray 
energy of 400~GeV} (GAMMA-400, see~\cite{gamma400}) is a Russian-led project 
for building a next generation gamma-ray telescope optimized for energies around $100$~GeV 
with the best possible angular and energy resolution and proton rejection factor~\cite{galper2013}.
Since the initial proposal that goes back to the late 1980ies the maximum target energy has
been raised, and the current GAMMA-400 design will be sensitive to gamma rays in the 
100 MeV -- 3 TeV energy range.

The GAMMA-400 detector is composed of
an anti-coincidence shield (AC),
a 10-layer double-sided silicon tungsten convertor tracker (C),
a time-of-flight system (TOF) made of two scintillation detectors (S1, S2),
two calorimeter detectors (CC1, CC2),
scintillation detectors (S3, S4) and a lateral detector (LD) surrounding the calorimeter,
and a neutron detector (ND) (cf.~Figure~\ref{fig:gamma400}).
The convertor tracker has a geometric area of $100\times100$~cm$^2$, a thickness of 1 radiation
length and a strip pitch of $80$~$\mu$m.
The imaging calorimeter (CC1) consists of 2 layers of double-sided silicon strip detectors with 
$80$~$\mu$m pitch, interleaved with CsI~(Tl) crystals.
The electromagnetic calorimeter (CC2) consists of CsI~(Tl) cubic crystals read out by
photodiodes.
The thickness of the calorimeter is 3 radiation lengths for CC1 and 22 radiation lengths for CC2,
which provides energy measurement capabilities with a resolution of about
$1\%$ (at 100 GeV).
High angular resolution ($\sim0.01^\circ$ for $>100$~GeV) is achieved by spatially separating C 
from CC1 by 50~cm at the expense of reducing the overall detection efficiency and field of view of 
the system.
The TOF system allows to identify backsplash particles in the AC, which together with a segmented
AC will reduce the self-veto rate.
The ND together with the calorimeter and other systems provides a proton rejection factor of $\sim10^6$,
allowing for an efficient particle background rejection.

GAMMA-400 is still in the design and early development phase.
The GAMMA-400 satellite has a total mass of 4.1 tons and a power consumption of 2~kW.
In addition to the gamma-ray telescope, the satellite will be equipped with the gamma-ray
bursts instrument Konus-FG that is composed of four NaI~(Tl) scintillation detectors used for 
burst localisation (operating in the $10-700$~keV energy range) and two NaI~(Tl) scintillation
detectors for spectroscopy (operating in the 10~keV -- 15~MeV energy range).
The satellite will be initially placed into a high-elliptical orbit with a perigee of 500~km, an apogee
of $300\,000$~km and an inclination of $51.4^\circ$, resulting in an orbital period of 7 days.
The orbit will then evolve into an approximately circular orbit with a median altitude of
$\sim150\,000$~km.
The launch with a Proton-M rocket is planned for $\sim2021$ for an expected mission duration
of $>7$~years.

\subsection{HARPO}

The {\it Hermetic ARgon POlarimeter} (HARPO, see~\cite{harpo})
is an instrument concept based on a time projection chamber (TPC) that will enable high angular
resolution and polarisation observations in the 3~MeV -- 3~GeV energy range~\cite{bernard2014}.
Conceptually, HARPO is similar to AdEPT in that it employs a large volume argon-based gas TPC
to track the electron-positron pairs induced by the incoming gamma rays.
A possible space mission would consist of 3 TPC layers, each layer consisting of 2 back-to-back
modules of volume $200\times200\times100$~cm$^3$, resulting in a gas sensitive volume
of 12~m$^3$.
The TPC will be filled with 100~kg of argon-based gas at a pressure of 5 bar.
A uniform electric field of the TPC will be established by a cubic field cage.
After drift, the ionisation electrons are amplified in a two-stage parallel plate avalanche chamber
composed of a micro-mesh on top of a microstrip detector with a 1~mm strip pitch.
The TPC will be enclosed in an anti-coincidence shield for particle background
reduction.

A demonstrator of the HARPO TPC composed of a $30\times30\times30$~cm$^3$
TPC filled with a $2-5$~bar argon-based gas mixture has been built~\cite{bernard2012}
and successfully tested in a 1.7 -- 74 MeV gamma-ray beam~\cite{wang2015}.
The concept for a potential space mission still needs to be defined.

\subsection{HERD}

The {\it High Energy cosmic-Radiation Detection} (HERD, see~\cite{herd}) facility is
one of several space astronomy payloads of the cosmic lighthouse program onboard China's Space 
Station, which is planned for  operation starting around 2020 for about 10 years~\cite{zhang2014}.
Similar to CALET and DAMPE, HERD will be a high-energy particle detector with gamma-ray
capabilities in the 100 MeV up to 10 TeV energy range.

HERD is composed of a 3-D cubic calorimeter (CALO) surrounded by microstrip silicon trackers 
(STKs) from five sides except the bottom.
CALO is a cubic calorimeter of size $63\times63\times63$~cm$^3$ which is made of nearly
10$^4$ pieces of granulated cerium-doped lutetium yttrium orthosilicate (LYSO) crystals of $3\times3\times3$~cm$^3$ each, providing
a thickness of 55 radiation lengths.
The top STK has a geometrical area of $70\times70$~cm$^2$ and is composed of 7 layers
of double-sided silicon strip detectors interleaved with tungsten.
The four side STKs have a geometrical area of $65\times50$~cm$^2$ and are in the baseline
design composed of 3 layers of double-sided silicon strip detectors.
A possible option consists of extending the side STKs to 7 layers and adding tungsten foils
to ensure a maximum field-of-view for gamma rays, which is one of the key features of the
HERD concept.
The entire system may be enclosed into plastic scintillators to reject low energy charged particle
background.
The total detector weight is estimated to $\sim2$~tons.

Some R\&D work is current ongoing to validate the key technologies of HERD.
The read out of 10$^4$ LYSO crystals with a dynamic range of $2\times10^6$ is particularly
challenging, and a system based on optical fibers and an image intensifier is currently
under study.
Building of a $1/20$ model of CALO is foreseen to evaluate the end-to-end performances of
the detector.

The HERD detector will have a total mass of 2 tons and a total power consumption of less
than 2 kW.

\subsection{PANGU}

The {\it PAir-productioN Gamma-ray Unit} (PANGU) mission is proposed as a candidate for the joint small
mission between the European Space Agency (ESA) and the Chinese Academy of Science (CAS).
PANGU targets the medium-energy range from 10 MeV to 1 GeV and is optimised for high angular
resolution~\cite{wu2014}.

The PANGU design is mainly driven by the stringent boundary conditions imposed by the
joint ESA and CAS call: a payload mass below 60~kg and a payload power budget below 60~W.
The PANGU collaboration attempts to fit into this budget by proposing an innovative design
composed of
a target-tracker system,
a magnetic spectrometer, and
an anti-coincidence detector.
The technology for the target-tracker system is not yet decided, and solutions employing 
thin silicon detectors or scintillating fibers are considered.
In the silicon option (PANGU-Si), the detector is composed of 50 layers composed of two single-sided
silicon strip detectors of 150~$\mu$m thickness.
In the fiber option (PANGU-Fi), the detector is composed of 50 detector modules, each including 
two orthogonal layers of fibers that are read out using silicon photomultipliers.
For both options, the tracker has a size of $50\times50\times30$~cm$^3$ and a thickness of 
0.16 radiation lengths.
To fit into the mass budget, PANGU will employ a magnetic spectrometer instead of a heavy
calorimeter, located below the target-tracker.
By measuring the magnetic deflection of the electron and positron tracks in the field of a permanent
magnet (with magnetic field in the range $0.05-0.2$ T), the moment of the particle will be inferred.
The spectrometer will contain six layers of silicon strip detectors to measure the track deflection.

PANGU will be either launched into a low inclination ($<5^\circ$) low Earth orbit (550~km altitude)
or to L2 with a nominal mission lifetime of $2-3$ years.
The satellite will be operated both in survey and pointing mode.

\section{Ground-based gamma-ray astronomy}
\label{sec:ground}

\subsection{The next steps}

\begin{table}[!t]
 \centering
  \begin{tabular}{l|C{2.5cm}|C{2.5cm}|C{2.5cm}|C{2.5cm}|C{2.5cm}}
   \hline \hline
   Parameter & CTA & HAWC & HiSCORE & LHAASO & MACE  \\
   \hline
   Site(s) &
     t.b.d. & 
     Sierra Negra (Mexico) & 
     Tunka Valley (Russia) & 
     Daocheng (China) & 
     Hanle (India) \\ 
   Altitude (m) &
     $\sim2\,000$ & 
     $4\,100$ & 
     $675$ & 
     $4\,300$  & 
     $4\,270$ \\ 
   Latitude &
     t.b.d. & 
     $19^\circ$~N & 
     $51.8^\circ$~N & 
     $29^\circ$~N & 
     $32.8^\circ$~N \\ 
    Start of operations &
     2020 & 
     2013 & 
     t.b.d. & 
     2020? & 
     2016 \\ 
    Lifetime (years) &
     30 & 
     10 & 
     t.b.d. & 
     $>10$ & 
     t.b.d. \\ 
    Energy range (TeV) &
     $0.02-300$ & 
     $0.1-100$ & 
     $50-10000$ & 
     $0.1-1000$ & 
     t.b.d. \\ 
    $\Delta E/E$ &
     $10\%$ & 
     $50\%$ & 
     $10\%$ & 
     $20\%$ & 
     t.b.d. \\ 
    A$_{\rm eff}$ (m$^2$) &
     $3\times10^6$ & 
     $30\,000$ & 
     $10^8$ & 
     $8\times10^5$ (KM2A) $10^6$ (WCDA) & 
     t.b.d. \\ 
    Sensitivity (mCrab) &
     $1$ & 
     $50$ & 
     $100$ & 
     $10$ & 
     t.b.d. \\ 
    Field of view & 
     $5^\circ-10^\circ$ & 
     $1.8$~sr & 
     $0.6$~sr & 
     $1.5$~sr & 
     $4^\circ$ \\ 
    Angular resolution &
     $0.05^\circ$ & 
     $0.5^\circ$ & 
     $0.1^\circ$ & 
     $0.3^\circ$ & 
     t.b.d. \\ 
   \hline
  \end{tabular}
 \caption{Summary of future observatories and experiments for ground-based gamma-ray astronomy.}
 \label{tab:ground}
\end{table}

Every ground-based gamma-ray telescope relies on the fact that a gamma ray interacting with
the molecules of the Earth atmosphere will produce an electromagnetic cascade of 
particles called an extensive air shower (EAS).
Ground-based gamma-ray telescopes distinguish into two broad classes:
those observing the Cherenkov light that is produced in the atmosphere by ultra-relativistic 
particles in an EAS, and
those observing the tails of the EAS when it reaches ground~\cite{GROUND}.
Imaging Air Cherenkov Telescopes (IACT) composed of up to several 100~m$^2$ large
optical mirrors and photo-multiplier tube based cameras with hundreds to thousands of pixels
have been proven most efficient to study gamma-ray induced atmospheric Cherenkov light, as 
they provide excellent angular resolution ($<0.1^\circ$) together with strong background rejection 
power ($>99\%$).
Drawbacks of IACTs are the relatively low duty cycles ($\sim10\%$) and narrow fields of view 
($\sim5^\circ$).
Well known examples of IACTs are the Whipple telescope, HEGRA, CAT and more recently,
H.E.S.S., VERITAS and MAGIC.
Water Cherenkov Detectors (WCDs) are the most successful devices for studying the EAS
particles when reaching ground.
Although they have poorer angular resolution and less background rejection power than
IACTs, their advantage lies in the excellent duty cycle ($>95\%$) and wide field of view 
($\sim2$~sr), making them ideal instruments for sky surveys, source monitoring and transient 
source discoveries.
Well known examples of WCDs are MILAGRO and more recently HAWC (cf.~section \ref{sec:hawc}).

Alternative techniques exist, yet so far have proven less successful.
Those include wavefront sampling of the atmospheric Cherenkov light employed by
CELESTE, HiSCORE (see section \ref{sec:hiscore}) and HAGAR (see section \ref{sec:mace}),
usage of scintillator detectors employed by Tibet AS$\gamma$ and LHAASO (see section \ref{sec:lhasso})
or resistive plate chambers used by ARGO-YBL for EAS detection on ground.
WCDs are also operated using the ``single particle technique''~\cite{vernetto2000} which searches
for a coincident increase of the detector rate in several detectors due to a transient phenomena, such 
as a gamma-ray bursts
(e.g., LAGO~\cite{sidelnik2014}, Auger~\cite{allard2007}), yet so far without any success.

Next generation IACTs will build on the experience gained with existing infrastructures, increasing
the sensitivity by covering a larger area on ground using more telescopes.
This will allow simultaneous observations of the same EAS with many telescopes, improving thus
the background rejection and angular resolution with respect to existing instruments.
By combining telescopes of different size classes, the energy range can be extended to
overlap at low energies with space-based instruments and to push high-energy detections
beyond 100~TeV.
Although photomultiplier tubes still present the most powerful sensors for detecting the faint
and short Cherenkov light flashes, Geiger-mode avalanche photodiodes, also known as SiPM, 
become increasingly interesting for usage in IACTs~\cite{bretz2014}.
This will enable IACT operations during moonlight conditions, increasing thus the available
observing time for a given infrastructure.

Next generation WCDs will also build on experience gained with past instruments (and specifically
with MILAGRO), covering larger surface areas to contain the sub-core of hadronic showers that is 
necessary to produce superior background rejection.
Furthermore, by moving the detector location to higher altitude, more shower secondaries will be
received, particularly at low energy, increasing thus substantially the instrument sensitivity and
lowering the energy threshold.
Also advances in the configuration of the detector, such as the optical isolation of the photomultiplier
tubes, will contribute to improve the instrument sensitivities.

Finally, with the maturity of the detection techniques and the increased scientific outcome of the
instruments, future ground-based instruments will move towards an open access model and
operations akin to other astronomical observatories.
This implies 
an increase of the instrument reliability (to keep operating costs manageable),
usage of astronomical standards and tools and provision of user support
(to reach also the non-expert community),
and the implementation of interoperable data archives (to assure the legacy of the observatory).

The following sections describe some of the proposed ground-based gamma-ray astronomy
facilities.
Table~\ref{tab:ground} provides summary information on the proposed instruments.
Details are given in the relevant sections.

\subsection{CTA}

\begin{figure}[!t]
  \centering
  \includegraphics[width=15cm]{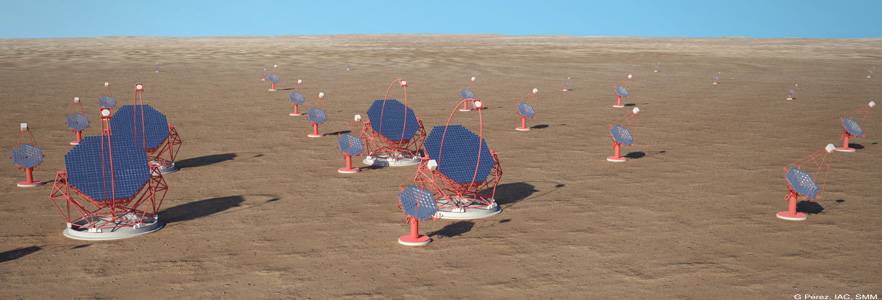}
  \caption{Artists view of a CTA array site}
  \label{fig:cta-artist}
\end{figure}

The {\it Cherenkov Telescope Array} (CTA, see~\cite{cta}) is a world-wide
project to create a large and sustainable IACT observatory that builds on the success and experience
gained from current imaging air Cherenkov telescopes (i.e.~H.E.S.S., MAGIC and 
VERITAS)~\cite{acharya2013}.
CTA will provide all-sky coverage by implementing two IACT arrays, one in the northern and one 
in the southern hemisphere, equipped in total with more than hundred IACTs of three different size 
classes to cover the energy range 20~GeV -- 300~TeV (cf.~Figure~\ref{fig:cta-artist}).
The project is currently developed by an international consortium comprising about $1\,200$
scientists and engineers from 28 countries, with construction starting in $2016-2017$ and
operations beginning in the 2020 time frame.
CTA will be operated as an open observatory, with a large fraction of the observing time being
made available in a proposal-driven guest observer program, and all data becoming public after
some limited proprietary period.

Three telescope size classes are required to cover the full CTA energy range in a cost-effective
way:
large sized telescopes (LSTs) for energies from the threshold to a few 100~GeV,
medium sized telescopes (MSTs) for the core energy range (100~GeV -- 10~TeV),
and small sized telescopes (SSTs) for high energies, above a few TeV.

The LST is an alt-azimuth telescope built as a tubular structure made of steel and carbon fiber 
reinforced polymer tubes.
The LST features a 23~m diameter parabolic dish equipped with 198 mirror facets, providing an
an effective mirror area of 369~m$^2$.
The telescope will have a focal length of $28$~m ($f/D=1.2$), and will be equipped with a PMT-based
camera comprising $2\,841$ pixels of $0.1^\circ$ angular diameter covering a field of view diameter
of $4.5^\circ$.
The camera will employ two gain chains for dynamic range coverage, and will sample pulses at a
nominal frequency of 1~GHz using dedicated ``Domino Ring Sampler'' chips.

The MST is a modified Davies-Cotton telescope with a reflector diameter
of 12~m and a focal length of 16~m ($f/D=1.3$) mounted on a polar mount.
The MST hosts 86 mirror tiles that provide an effective mirror area of $88$~m$^2$.
The telescope will be equipped with a PMT-based camera for which two options exist:
NectarCAM, a modular $1\,855$ pixel camera employing two gain chains for dynamic range
coverage and that samples pulses with a nominal frequency of 1~GHz using the
dedicated NeCTAr chip, and
FlashCam, a $1\,764$ pixel camera employing a non-linear pulse amplifier to cover the
dynamic range and that continuously samples the PMT signal at 250~MHz frequency using
commercial FlashADCs.
The pixel angular diameter of both cameras is $0.18^\circ$ covering a field of view diameter
of $8^\circ$ for NectarCAM and $7.7^\circ$ for FlashCam.

A second design exists for the medium sized telescopes that is based on a two mirror optical
system in Schwarzschild-Couder configuration (SCT).
It is planned that the SCTs will complement the MSTs in the southern array as an extension.
The SCT features a segmented 10~m diameter primary and a segmented $5.4$~m diameter
secondary mirror that provide an effective light collecting area of $50$~m$^2$ for a focal
length of $5.6$~m ($f/D=0.6$).
The telescope will be equipped with a camera comprising $11\,328$ pixels of $0.07^\circ$
angular diameter covering a field of view diameter of $8^\circ$.
The camera can accommodate either SiPM (baseline) or multi-anode PMTs as photosensors, the
readout is performed using the dedicated TARGET chip that samples pulses with a nominal 
frequency of 1~GHz.

For the SST, three design options exist:
SST-1M which is based on a single reflection Davies-Cotton design, and
ASTRI and GCT, which are both based on double reflection Schwarzschild-Couder designs.

The SST-1M is an alt-azimuth telescope with a 4~m reflector diameter and a focal length of 
$5.6$~m ($f/D=1.4$).
The mirror is composed of 18 tiles that provide an effective mirror area of $6.5$~m$^2$.
The telescope is equipped with a SiPM-based $1\,296$ pixels camera that continuously
samples the PMT signal at 250~MHz frequency using commercial FlashADCs.
The camera will feature specifically developed large area hexagonal SiPM detectors
that are gain stabilised using a temperature -- bias voltage feedback loop.
The pixel angular diameter is $0.24^\circ$ covering a field of view diameter of $9^\circ$.

The ASTRI telescope adopts an alt-azimuth design featuring a segmented $4$~m diameter
primary and a monolithic $1.8$~m diameter secondary mirror that provide an effective light 
collecting area of $6$~m$^2$ for a focal length of $2.2$~m ($f/D=0.5$).
The telescope will be equipped with a SiPM-based $1\,984$ pixel camera with 
$0.17^\circ$ pixel angular diameter covering a field of view diameter of $9.6^\circ$.
The photosensors are readout using the dedicated CITIROC chip providing trigger and charge
measurement informations, as well as SiPM gain adjustment (no pulse shape sampling is 
foreseen).

GCT is a alt-azimuth dual mirror Schwarzschild-Couder telescope with a segmented 4~m
diameter primary and a segmented or monolithic 2~m secondary mirror that provide an 
effective light collecting area of 8.8~m$^2$ for a focal length of $2.3$~m ($f/D=0.6$).
The telescope will be equipped with a camera comprising $2\,048$ pixels of $0.17^\circ$
angular diameter covering a field of view of about $9^\circ$ in diameter.
Similar to the SCT camera, the GCT camera can accommodate either multi-anode PMTs or
SiPM as photosensors that are read out at 1~GHz frequency using the TARGET chip.

The baseline array layouts comprise
4 LSTs, 25 MSTs, 28 SCTs and 70 SSTs in an area of radius 1.5~km at the southern site, and
4 LSTs and 15 MSTs within an area of radius $0.4$~km at the northern site.
Observing modes include 
the full array pointing towards a single source,
sub-arrays targeting each a specific source, and
divergent telescope pointing with overlapping fields of view to cover a large field of the
sky in survey mode.
Cameras will trigger independently of other telescopes to suppress night-sky background light.
Additional inter-telescope triggers and array trigger schemes are under consideration to select
shower signatures, to stabilize the readout rates and to recognize background events 
(e.g. muon rings, obvious hadrons).
Off-line background reduction will rely on well proven techniques that rely on differences in shower
image morphology between gamma-ray and hadron-induced showers.

CTA will be a factor of 10 more sensitive than any existing VHE instrument, reaching 1~mCrab
for a typical observing time of 50~h.
CTA will cover four orders of magnitude in energy, from a few tens of GeV to a few
hundred TeV, again a factor 10 more than any existing facility.
By selecting a subset of gamma-ray induced cascades detected simultaneously by many
of its telescopes, CTA can reach angular resolutions of better than 2 arcminutes for energies
$>1$~TeV.
In its core observing program, CTA will perform deep surveys of the Galactic plane and the
extragalactic space down to a uniform sensitivity of a few mCrab, providing an unprecedented
census of VHE source population in the Universe.
The expected lifetime of the observatory is 30 years.

\subsection{HAWC}
\label{sec:hawc}

\begin{figure}[!t]
  \centering
  \includegraphics[width=15cm]{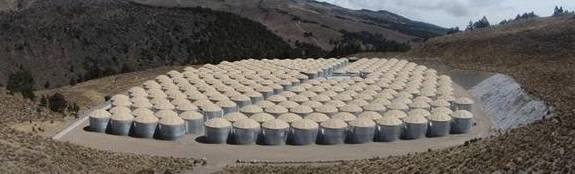}
  \caption{The High-Altitude Water Cherenkov Observatory (HAWC) in Sierra Negra, Mexico.}
  \label{fig:hawc}
\end{figure}

The {\it High-Altitude Water Cherenkov} observatory (HAWC, see~\cite{hawc})
is a direct successor to the MILAGRO instrument which has demonstrated that a water Cherenkov 
detector (WCD) optimized for reconstructing gamma-ray air showers can provide observations of
TeV gamma-rays with a wide field of view and high duty cycle.
HAWC is essentially a re-deployment of the MILAGRO photo-multiplier tubes (PMTs) and
electronics in a different configuration at an altitude above $4\,000$~m, which results
in a lower energy threshold (down to $\sim100$~GeV) and a 15-fold increase in gamma-ray
sensitivity~\cite{lauer2014}.

HAWC, which is fully deployed since beginning of 2015, consists of 300 cylindrical WCDs
of 7.3~m diameter and 4.5~m height, filled with clear water and instrumented each with four 
upward facing PMTs~\cite{goodman2013}.
The tanks are deployed closely packed over $20\,000$~m$^2$ on a $4\,100$~m plateau near
the Sierra Negra, Mexico.
Figure~\ref{fig:hawc} shows a picture of the fully deployed instrument.

To separate gamma-rays from the dominant hadronic background, a gamma/hadron filter 
is applied during data analysis that is based on the compactness of the shower footprint.
Hadronic showers show an increased granularity due to the higher probability of spatially 
separated secondary showers.
Above a few TeV, the hadron rejection efficiency is better than $10^2$ while retaining at the 
same time about $50\%$ of the gamma-ray induced showers~\cite{tepe2012}.

HAWC has an aperture of $<45^\circ$ in zenith angle resulting in an instantaneous field of view 
of about $1.8$~sr.
Given the location of the observatory at $19^\circ$ northern latitude, HAWC provides a
daily sky coverage of 2/3 of the whole sky ($\sim8$~sr).
Any source in the covered sky region is typically seen for 4~hours per day.
Within one year of observations, HAWC will reach a source detection sensitivity above
$>2$~TeV of $\sim50$~mCrab for a good fraction of the observable sky~\cite{goodman2013}.
The expected lifetime of the observatory is 10 years.

\subsection{HiSCORE}
\label{sec:hiscore}

The {\it Hundred Square-km Cosmic ORigin Explorer} (HiSCORE) is a 
large-area (up to 100~km$^2$),
wide-angle (field of view $\sim0.6$ sr)
air-shower detector concept aiming to explore cosmic rays and gamma rays in the
energy range from a few 10s of TeV to above 1 PeV using the non-imaging air-Cherenkov
detection technique~\cite{tluzykont2014}.
In its standard array layout,  HiSCORE is composed of 484 detector stations distributed in a regular 
grid over a surface area of 10 km$^2$ with an inter-station spacing of 150~m.
Each HiSCORE station consists of four PMTs, each equipped with a light-collecting Winston
cone of $30^\circ$ half-opening angle pointing to the zenith
(a tilted mode is also envisioned to increase the accessible sky area).
The photon arrival directions are reconstructed using an analytical model for the arrival time of
Cherenkov photons at the detector stations.
The photon energies are reconstructed from the measured light densities.
Particle background is rejected on the basis of the reconstructed energy and shower heights,
with a quality factor (retained photon events over square root of retained particle events)
reaching about 2 at 1 PeV.
In a final array configuration covering 100 km$^2$, HiSCORE may reach a sensitivity of about
100 mCrab at 1~PeV ($10^{15}$~eV) after 5 years of survey observations (corresponding to 
1000 hours of exposure time)~\cite{tluzykont2014}.

An engineering array covering a surface of $\sim1$~km$^2$ is currently deployed in the Tunka valley
(Russia), forming part of a larger infrastructure named the
{\it Tunka Advanced Instrument for cosmic ray physics and Gamma Astronomy} (TAIGA)~\cite{taiga}.
TAIGA comprises also the Tunka-133 array~\cite{antokhonov2011}
and should furthermore host up to 16 imaging air Cherenkov telescopes and a net of surface and
underground stations for measuring the muon component of air showers~\cite{budnev2014}.
In a next stage it is planned to extend the HiSCORE array to a surface area of 10~km$^2$.
Ultimately, a configuration covering 100~km$^2$ is targeted which would provide interesting
science potential in the $>100$~TeV gamma-ray domain.

\subsection{LHAASO}
\label{sec:lhasso}

The {\it Large High Altitude Air Shower Observatory} (LHAASO) is a planned experiment for 
gamma-ray and cosmic-ray physics that will be located at Daocheng (China) at an altitude of 
$4\,300$~m.
LHAASO will be a hybrid detector array composed of 
a 1~km$^2$ large detector array (KM2A),
a $90\,000$~m$^2$ large water Cherenkov detector array (WCDA),
24 wide field Cherenkov telescopes (WFCTA), and
a $5\,000$~m$^2$ large array of shower core detectors (SCDA).
While WCDA and KM2A aim at the detection of gamma rays in the $0.1-1000$~TeV energy
range, WFCTA and SCDA target the study of cosmic rays above 30~TeV.

WCDA covers the $0.1-30$~TeV energy range and is composed of four neighbouring 
$150\times150$~m$^2$ pools, each of which is partitioned by black curtains into 
$5\times5$~m$^2$ cells with an effective water depth of 4~m.
Each cell is viewed by a hemispherical 8'' photo-multiplier tube at the bottom that faces
upward to receive the Cherenkov light produced in the water tank.
Cosmic-ray rejection is achieved by measuring the compactness of the shower
(gamma-ray induced showers lead to a more compact signal than proton-induced
showers).
Cosmic-ray rejection efficiencies of $10^3$ or better are predicted above a few TeV.

KM2A covers the $30-1000$~TeV energy range and is composed of $5\,635$ electromagnetic
particle detectors (EDs) and $1\,221$ muon detectors (MDs)~\cite{liu2013}.
Each ED consists of a 2~cm thick square plastic scintillator with an area of $1\times1$~m$^2$,
covered by a $0.5$~cm thick lead plate that serves as pair convertor for incoming gamma rays.
The scintillator is read out through a set of wavelength shifting fibers using a photo-multiplier 
tube.
Each MD consists of a cylindrical water tank with diameter of $6.8$~m and height of
$1.2$~m, equipped with an inner Tyvek coating and viewed by a single 8'' or 9'' photo-multiplier 
tube.
Each MD is covered by an overburden of $2.5$~m soil, which results in a muon energy threshold
of $1.3$~GeV to mask electromagnetic particles in showers.
The EDs and MDs are located at the corners of regular triangles with side lengths of
15~m and 30~m, respectively.
Gamma/hadron separation is achieved by comparing the number of muons
measured in the MDs to the number of electrons measured in the EDs.
While gamma-ray induced showers are electron rich, proton-induced showers are muon
rich.
Cosmic-ray rejection efficiencies of $10^5$ or better are expected above 50 TeV~\cite{cui2014}.

WCDA and KM2A have an aperture of $<40^\circ$ in zenith angle resulting in an
instantaneous field of view of $1.5$~sr.
Given the location of the observatory at $29^\circ$ northern latitude, LHAASO provides thus a
daily sky coverage of $7$~sr, corresponding to roughly half of the sky.
During one day, most of the sources are visible for $4-6$ hours, depending on declination.
Specifically, the Galactic plane is visible in the longitude range from $20^\circ$ to $225^\circ$.

Engineering arrays at scales of $1\%-10\%$ of the full project have been built up at the 
ARGO-YBL site in Tibet.
LHAASO has been included in the Chinese roadmap of infrastructure construction for basic
science in a short term (5 years).
The experiment may become operational in the 2020 time-frame.

\subsection{MACE}
\label{sec:mace}

The {\it Major Atmospheric Cherenkov Experiment} (MACE) telescope is a 21~m diameter
imaging air Cherenkov telescope that will be operational by 2016 at the
{\it Himalayan Gamma Ray Observatory} (HiGRO) at Hanle located in Northern India at an
altitude of $4\,270$~m~\cite{koul2011}.
HiGRO is currently equipped with the {\it High Altitude GAmma Ray} (HAGAR) telescope array
which has been operating since 2008.
HAGAR employs the wavefront sampling technique to reconstruct the arrival direction of the
incoming gamma rays, yet so far, only the brightest gamma-ray sources (i.e.~Crab, Mkn~421)
have been convincingly detected~\cite{britto2012}.

MACE will be composed of a 150~tons mechanical steel structure with an overall height of
45~m.
The 21~m diameter basket will form a paraboloid mirror shape with a focal length
of 25~m.
The mirror will be composed of 356 diamond turned spherical aluminum honeycomb 
panels that will provide a total reflective area of $\sim330$~m$^2$.
MACE will be equipped with a PMT-based modular camera comprising 1088 pixels.
The camera will have a pixel resolution of $0.125^\circ$ and a field of view of $4^\circ$.

The manufacturing of the structural elements of the telescope has been completed and the 
proof assembly of the mechanical structure along with the drive servo system has been 
completed near the city of Hyderabad, South India.
The mechanical structure is in the process of being dismantled and shipped to Hanle.

Simulations suggest that MACE will reach an energy threshold of 20~GeV\cite{koul2011}.

\section{Conclusions}
\label{sec:conclusions}

\begin{figure}[!t]
  \centering
  \includegraphics[width=16cm]{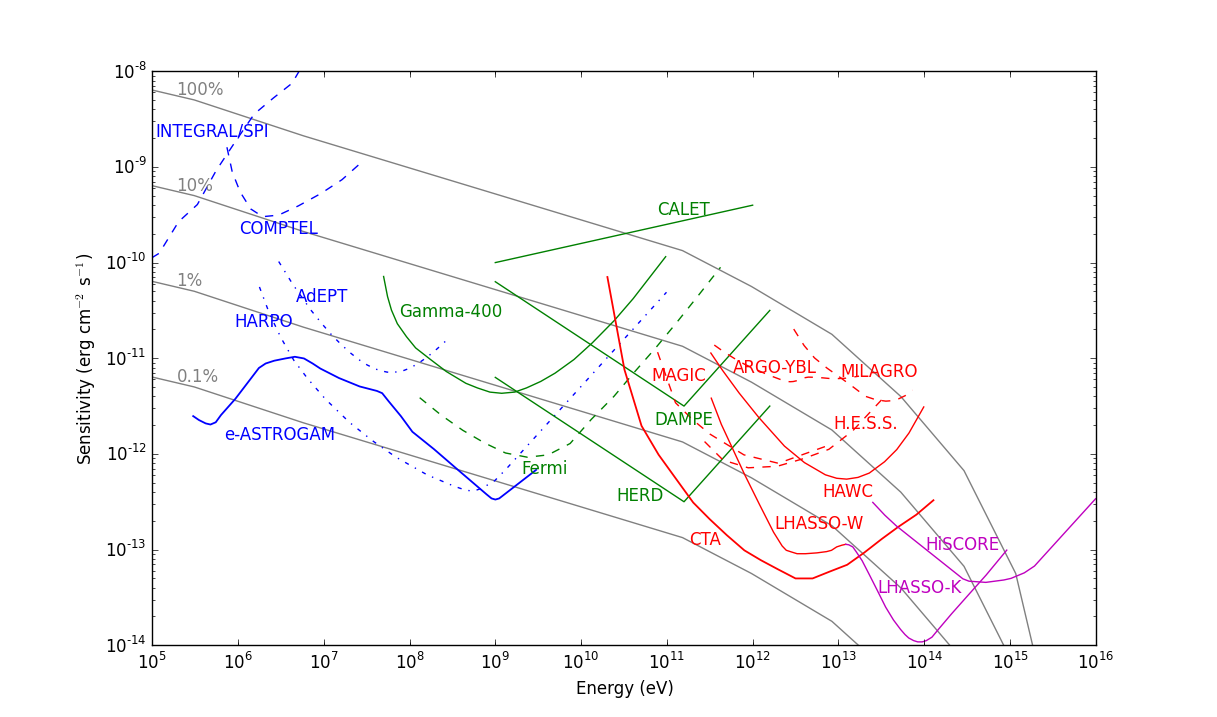}
  \caption{
  Differential $5\sigma$ sensitivity of gamma-ray telescopes
  (1 erg cm$^{-2}$ s$^{-1}$ = 1 mW m$^{-2}$).
  Future instruments are shown as solid lines, concepts based on R\&D work are shown as
  dashed-dotted lines and existing or past instruments are shown as dashed lines.
  Colors distinguish the energy domains (blue for low and medium energy, green for high energy,
  red for very high energy and magenta for ultra high energy.
  The grey lines show the Crab differential energy flux, as well as 10\%, 1\% and 0.1\% of that
  flux for reference.
  The lowest of the lines corresponds to a differential energy flux of 1 mCrab.
  %
  The INTEGRAL sensitivity is for a typical observing time of $10^6$ s~\cite{roques2003}.
  %
  The COMPTEL sensitivity is for the observing time accumulated during the whole duration
  of the CGRO mission ($\sim9$ years).
  %
  The AdEPT sensitivity is for a typical observing time of $10^6$ s~\cite{hunter2014}.
  %
  The
  e-ASTROGAM~\cite{tatischeff2015},
  HARPO~\cite{bernard2013},
  and Fermi-LAT~\cite{ackermann2012b}
  sensitivities are given after 3 years of scanning observations for a source at high Galactic 
  latitudes~\cite{tatischeff2015}.
  %
  The GAMMA-400 sensitivity is for a 30 days long observation of a source at high Galactic
  latitude~\cite{sparvoli2014}.
  %
  The CALET, DAMPE and HERD sensitivities are for 1 year of observation time (adapted 
  from~\cite{zhang2014}).
  %
  The 
  MAGIC stereo system~\cite{aleksic2012},
  H.E.S.S.,
  and CTA~\cite{bernloehr2013}
  sensitivities are for an effective exposure time of 50 hours.
  %
  The Argo-YBL sensitivity is after 6 years of observations~\cite{disciascio2014}.
  %
  The MILAGRO sensitivity is for 1 year of observations~\cite{goodman2013}.
  %
  The HAWC sensitivity is after 5 years of observations~\cite{goodman2013}.
  %
  The HiSCORE sensitivity is for a 100 km$^2$ array after 5 years of survey observations,
  equivalent to an on-source exposure time of 1000 hours~\cite{tluzykont2014}.
  %
  The LHASSO sensitivity is after 1 year of observations~\cite{vernetto2014}.
  %
  }
  \label{fig:sensitivities}
\end{figure}

Figure~\ref{fig:sensitivities} summarizes the differential $5\sigma$ point source sensitivities
as function of energy for those instruments and mission concepts for which this information was
available.
For INTEGRAL and AdEPT, sensitivities have been quoted in the literature for a confidence level 
of $3\sigma$, and these have been converted to a confidence level of $5\sigma$ by multiplying 
the quoted values with $5/3$.
Note that the assumed effective observing times differ between the various instruments.
For instruments with a narrow field of view, sensitivities are shown for typical source exposure
times (e.g.~INTEGRAL, GAMMA-400, MAGIC, H.E.S.S., CTA).
For wide field of view instruments, the typical sensitivity after several years of sky survey
operations are shown (e.g.~COMPTEL, e-ASTROGAM, HARPO, CALET, DAMPE, HERD,
ARGO-YBL, MILAGRO, HAWC, HiSCORE, LHASSO).
The exception to this rule is AdEPT for which the sensitivity for an exposure time of $10^6$ s
was taken from~\cite{hunter2014}.
In the background-dominated regime (which essentially applies to AdEPT),
sensitivities scale with $1/\sqrt{t}$, hence for a continuous one year long observation,
the sensitivity would improve by a factor of $\sim6$, making AdEPT equivalent to HARPO
(which is not surprising because both instruments are very similar).
For the source-dominated region the sensitivity improves with $1/t$.~\footnote{The source-dominated
region is typically located at the high-energy end for the high- and very-high-energy domains; 
the low- to medium-energy domains are always background dominated.}
For instruments in the high-energy domain, where the diffuse Galactic emission is dominating the
background for point source studies, sensitivities are quoted for sources at high Galactic latitude
(i.e. where the background is low).
Generally, the sensitivity degradation towards the Galactic plane due to diffuse emission depends 
on the angular resolution of the instrument.
For Fermi-LAT, for example, the sensitivity near the Galactic centre is by a factor of 10 worse
than at high latitudes.
Owing to the better angular resolution, this penalty should be smaller for Gamma-400.

Figure~\ref{fig:sensitivities} illustrates that a substantial improvement in sensitivity can be achieved
in the medium-energy domain by a mission like e-ASTROGAM.
At MeV energies, this mission would be $10-30$ times more sensitive than its predecessor
COMPTEL, and even below 1~MeV, a dramatic sensitivity increase would be achieved
with respect to INTEGRAL/SPI (by about a factor 100).
e-ASTROGAM will reach up to GeV energies, with a sensitivity comparable to that of Fermi-LAT,
and will cover the still poorly explored energy range around 100~MeV where the Fermi-LAT
sensitivity drops rapidly.
This is a crucial energy range, as it features the turn-over of hadronic particle spectra, and thus
is a unique carrier of distinctive information about the nature of the accelerated particles.
e-ASTROGAM would also cover the energy range that is crucial to explore the intensity and
physics of the still elusive low-energy cosmic-ray component.
e-ASTROGAM furthermore covers the MeV domain that is rich of gamma-ray lines which give access
to nuclear cosmic-ray excitations, nucleosynthesis processes, and the positron-electron
annihilation features below $\sim511$~keV.
The AdEPT and HARPO concepts are also promising in the medium- to high-energy domains,
though both employ large volume gaseous time projection chambers that still need to demonstrate
their compliance with a space environment.
Stratospheric balloon flights, as planned for example for the AdEPT instrument, are here crucial
to test the reliability of the instrument in a space-representative environment.

Another important leap in sensitivity will be achieved in the very high energy domain by CTA,
with a factor of $\sim10$ increase with respect to current telescopes.
One of the key features of CTA will be the extended energy range from a few tens of GeV to above
100 TeV, which is nicely visible in Fig.~\ref{fig:sensitivities}.
With a few 100 hours exposure time, CTA will reach milli-Crab differential sensitivity, making it
the most sensitive gamma-ray telescope ever.
CTA will reach the expected gamma-ray emission from annihilation of WIMP particles with masses 
in the few 100 GeV to few TeV domain, provided that their annihilation cross section corresponds 
to the thermal relic density value.
CTA will reveal thousands of very high energy gamma-ray sources, enabling comprehensive
population studies of particle accelerators in the Universe.
For example, CTA will detect gamma-ray emission from young supernova remnants located at
the opposite edge of the Galaxy, accessing thus the entire Galactic volume to search for the
still elusive Galactic PeVatrons.
For steady sources, CTA will surpass the Fermi-LAT sensitivity above $\sim50$~GeV.
For short-time phenomena, such as gamma-ray bursts or gamma-ray flares, CTA will be several 
orders of magnitude more sensitive than Fermi-LAT even at lower energies, owing to its
huge effective detection area of $>10^6$~m$^2$ (compared to $\sim1$~m$^2$ for Fermi-LAT).
CTA will thus open a new window for time-domain astronomy, probing for example sub-minute
variability in active galactic nuclei.

LHASSO will explore even higher energies than CTA, thanks to the KM2A detector that measures 
the muon content in atmospheric cascades.
Also the HiSCORE concept, once deployed over an area of 100 km$^2$, provides interesting
performances in the ultra-high-energy domain.
At lower energies, the WCDA detector of LHASSO should be more sensitive than HAWC but will
still be less sensitive than CTA.
Yet the large field of view and the high duty cycle provide complementary performances to CTA
that are well suited for conducting sky surveys and generating transient source alerts.

To conclude, the instrumental prospects for gamma-ray astronomy are rich, and concepts for
new instruments exist throughout the entire energy domain.
HAWC  just went online, the next to come will be the CTA observatory in the 2020 time frame.
LHASSO is also scheduled to become operative by $\sim2020$.
e-ASTROGAM may be launched by 2029, provided that it will be selected by ESA as M5
mission.
It is noteworthy that among all these instruments, only e-ASTROGAM and CTA will operate as open
astronomical observatories, hence one may speculate that their scientific impact will be
largest.
In any case, given the scientific challenges, and given the rich instrumentation that is currently
planned and developed, the future of gamma-ray astronomy promises to be exciting.


\end{document}